\documentclass[12pt, a4paper]{article}
\usepackage{amsmath}
\usepackage{graphicx}
\usepackage{hyperref}

\begin{document}
\title{On  the Relativistic Quantum Plasma}
\author{Rashid Ahmad${}^{\star}$, Ikramullah, Saqib Sharif, Shakir Husain,\\ Fida Younus Khattak
\\ Department of Physics\\Kohat University of Science and Technology \\ 26000  Kohat, Pakistan \\
 ${}^{\star}$rashidahmad@kust.edu.pk }

\maketitle
\begin{abstract}
 Recently the interest in relativistic quantum plasma is increasing primarily to understand the fundamentals of the plasma behaviour and its properties. Mathematical models used to investigate these plasma are still need to be matured. Especially, the relativistic quantum electron-ion plasma are modeled using the Klein-Gordon equation and the Dirac equation for  relativistic electrons. However, different properties of these plasma are investigated without anti-particles.  We note that in order to preserve causality   relativistic quantum plasma must contain anti-particles for  relativistically dynamical components of  the  plasma .
\end{abstract}

\section{Introduction}
Recently there have been some articles using the Klein-Gordon and the Dirac equations to study different properties of the quantum electron-ion plasma in  relativistic  regime. In some of such studies electrons in the plasma are modeled with the Klein-Gordon equation obviously neglecting spin and hence the degeneracy pressure effects \cite{Bengt},\cite{Padma},\cite{Haas} and in some of them electrons are modeled  with the Dirac equation \cite{Eliasson}. However, the most important aspect of the Klein-Gordon equation and the Dirac equation neglected is the negative energy solution.\\
\\\hspace*{15 mm}   It is well known that  the quantity $\rho=\psi^{\star}\frac{\partial \psi}{\partial t}-\tilde{\psi}\frac{\partial \psi^{\star}}{\partial t}$ obtained from the Klein-Gordon equation is not positive definite and  should be interpreted as the charge density rather than probability density. Both positive and negative energy solutions are indispensable. The sign change of $\rho$ simply shows that two solutions describe  states of opposite charges. Therefore,   procedures developed in \cite{Bengt},\cite{Padma},\cite{Haas} are not valid  without inclusion of anti-particles for dynamical components of the plasma.\\
\\\hspace*{15 mm} On the other hand  the quantity $\psi\psi^{\dag}=\mid\psi\mid^2$ obtained from the Dirac equation is positive definite. However, to preserve  causality  positrons must also be included in the relativistic quantum electron-ion plasma considered in \cite{Eliasson} where in order to include spin effects the  Dirac equation is used for electrons.\\
\\\hspace*{15 mm} However, in this article we only  analyze the procedure developed in  \cite{Bengt} incorporating  negative energy states. We also provide an explanation for   dispersion curves obtained in \cite{Bengt}.  \\
\\\hspace*{15 mm} This article is organized as follows. In section \ref{causality and anit-partciles} we show that any relativistic quantum mechanical theory must contain anti-particles for  corresponding relativistically dynamical particles. In section \ref{Field Viewpoint} we emphasize the necessity of field description of the relativistic electrons in the relativistic quantum mechanical plasma.    In section \ref{klein-gordon-poisson model} we give the mathematical model for calculating the dispersion relation for electrostatic waves in electron-positron-ion three component plasma and in section \ref{Dispersion Relation} we derive the dispersion relation for the quantum electron-positron-ion plasma in relativistic regime .
\section{Causality and Anti-particles}\label{causality and anit-partciles}
In \cite{WEINBERG} Steven Weinberg demonstrated beautifully (page 61-63) that in the relativistic quantum mechanics without anti-particles causality is violated . However, it is important to note that this is not true in the non-relativistic quantum mechanics or in the relativistic classical mechanics. For more mathematical proof see \cite{Streater} and  especially for similar discussion in terms of scalar fields associated with the Klein-Gordon equation see \cite{0201503972}.  \\
\\\hspace*{15 mm} In quantum plasma at  wavelengths comparable with the Compton wavelength electrons become relativistic and due to the uncertainty principle of the quantum mechanics it is possible for  electrons to tunnel into space-like regions.
 In the special theory of relativity  for space-like intervals temporal order of events is observer dependent i.e.  two lorentz observers generally do not agree on the order of events and hence causality is violated.  \\
\\\hspace*{15 mm} For instance, in the relativistic quantum plasma relativistic electrons in some inertial frame of reference say $S$ can start at spacetime point $x_1$ at time $t_1$ and end at spacetime point $x_2$ at latter time $t_2$ then the  spacetime interval is allowed by uncertainty principle to be space-like. i.e.
\begin{align}
(x_2-x_1)^2-(t_2-t_1)^2<0
\end{align}
The probability of a particle reaching $x_2$ if it starts at $x_1$ is non-negligible as long as
 \begin{align}
 (x_2-x_1)^2-(t_2-t_1)^2\leq \frac{\hbar^2}{m^2}
 \end{align}
 where $m$ is the mass of the electron and  $\hbar$ is the plank's constant divided by $2\pi$. However, we can find another inertial frame of reference say $S'$ where the electrons reach at spacetime point $x_2$ before they started  at spacetime point $x_1$ which violates causality and is logically impossible . The only way out of this  logical paradox is that the $S'$ must see the different particle at $x_2$. Since the mass is the lorentz invariant both particles should be of the the same mass but of the opposite charge. \\
\\\hspace*{15 mm}This is clear from the above  example that  electron-ion  plasma does not exist in relativistic quantum regime. Only those plasma where anti-particles for the relativistically dynamical component of the plasma are present such as positrons for electrons can exist in nature and do not violate causality and hence are allowed by the special theory of relativity.\\
\\\hspace*{15 mm}This third component of the plasma then effects the dispersion relation of the electrostatic waves in the plasma and all other properties of such plasma are different and need to analysed.
\section{Field Viewpoint}\label{Field Viewpoint}
 The Klein-Gordon and the Dirac equations using the Feshbah-Villars formalism can be  split into two coupled Schrodinger and Pauli like single particle relativistic equations one each for electron and positron \cite{Villars}. Victor Kowalenko, Norman E. Frankel and Kenneth C. Hines used linear response theory and these equations to derive dispersion relation for two component electron-positron plasma \cite{Victor}.
 However,  separation of charge degrees of freedom of both the Klein-Gordon and the Dirac equations is only possible  in the presence of weak external electrostatic field.\\
\\\hspace*{15 mm} For more general situations we have to stick to the multi-particle nature of the plasma and for this we have to use the quantum field theoretical descriptions of the Klein-Gordon and the Dirac equations \cite{Quang},\cite{Mark}. \\
\\\hspace*{15 mm} We have to treat the solution of the Klein-Gordon and the Dirac equations as the quantum mechanical field operator rather than as a wavefunction. In  literature sometimes this procedure is called second quantization.

\section{The Klein-Gordon-Poisson Model}\label{klein-gordon-poisson model}
For the electron-positron-ion plasma in the relativistic quantum mechanical  regime with stationary  ion background and relativistically dynamical electrons and positrons the Klein-Gordon-Poisson model can be used to calculate the  dispersion relation for electrostatic waves. Such a system can be generalized to study (relativistic) electromagnetic wave interaction with the  relativistic quantum mechanical plasma and other properties of such plasma by including Maxwell equations in the system.\\
\\\hspace*{15 mm} The Klein-Gordon equation coupled to the electromagnetic scalar and vector potentials
\begin{align}
\label{kleingordon}\mathcal{W}^2\psi-c^2\mathcal{P}^2\psi-m^2c^4\psi=0
\end{align}
where
\begin{align}
\label{timederivative} \mathcal{W}=i\hbar\frac{\partial}{\partial t}+e\phi(r,t)
\end{align}
 and \footnote{We have used the  notations of \cite{Bengt} for the sake of convenience.}
 \begin{align}
\label{spacederivative} \mathcal{P}=-i\hbar\nabla+eA(r,t)
 \end{align}
describes the complex scalar field theory. Where $\psi$ creates  positively charged  particles and destroys negatively charged  particles . We also have   $\psi^{\dag}$ which does the opposite operations.
\\\hspace*{15 mm}
Supposing the singly charged ions we impose the charge neutrality condition on this three component plasma
\begin{align}
n_{0e}=&n_{0p}+n_{0i}
\end{align}
where
\begin{align}
n_{0e}-n_{0p}=&n_{0i}
\\n_{0effective}=&n_{0i}
\end{align}
and the effective charge density could be obtained from the Klein-Gordon equation
\begin{align}
\label{chargedesnity}\rho_{effective}=&-\frac{ e}{2m_ec^2}\Bigg(\psi^{\dag}\mathcal{W} \psi+\psi(\mathcal{W} \psi^{\dag})\Bigg)
\end{align}
The charge density is then the difference of two positive definite charge densities
\begin{align}
\rho_{effective}=&\rho_{electrons}-\rho_{positrons}
\end{align}
The Poisson equation for the effective charge density in the presence of neutralizing ion background
\begin{align}
\nabla^2\phi=-\frac{1}{\epsilon_0}(\rho_{effective}+\rho_i)
\end{align}
completes the model.
\section{The Dispersion Relation}\label{Dispersion Relation}
To find the dispersion relation for electrostatic oscillations in the quantum relativistic  electron-positron-ion plasma, the Klein-Gordon equation (\ref{kleingordon}) can be put into a more useful form using the transformation $\psi=\tilde{\psi}(r,t)\text{exp}(\frac{-imc^2t}{\hbar})$
\begin{align}
\label{transformedkleingordon}\Big(i\hbar\frac{\partial}{\partial t}+m_ec^2+e\phi\Big)^2\tilde{\psi}(r,t)+c^2\hbar^2\nabla^2\tilde{\psi}(r,t)-m_e^2c^4\tilde{\psi}(r,t)=0
\end{align}
and correspondingly the charge density for the effective electrons takes the following form for the equation (\ref{transformedkleingordon})
\begin{align}
\label{modfiedchargedensity}\rho_{effective}=-\frac{i\hbar e}{2mc^2}\Bigg(\tilde{\psi}^{\dag}\frac{\partial \psi}{\partial t}-\tilde{\psi}\frac{\partial \psi^{\dag}}{\partial t}\Bigg)-\Bigg(1+\frac{e\phi}{mc^2}\Bigg)e|\tilde{\psi}|^2
\end{align}
 We linearize the  equations (\ref{transformedkleingordon}) and (\ref{modfiedchargedensity}) with $\phi(x,t)=\phi_1(x,t)$ with $\phi_0=0$ and $\tilde{\psi}(x,t)=\tilde{\psi}_0+\tilde{\psi}_1$ to obtain
\begin{align}
\label{modfiedkge4}\nonumber&\hbar^2\Bigg(-\frac{\partial^2}{\partial t^2}+c^2\nabla^2\Bigg)\tilde{\psi}_1(r,t)+\Bigg(2m_ec^2e\phi_1(r,t)+i\hbar e\frac{\partial}{\partial t}\phi_1(r,t)\Bigg)\tilde{\psi}_0
\\&+2i\hbar m_ec^2\frac{\partial}{\partial t}\tilde{\psi}_1(r,t)=0
\end{align}
and
\begin{align}
\nonumber\label{modfiedkge7}\rho_{effective}+\rho_i=&-\frac{i\hbar e}{2mc^2}\Bigg(\tilde{\psi}_{0}^{\dag}\frac{\partial \psi_{1}}{\partial t}-\tilde{\psi}_{0}\frac{\partial \tilde{\psi}_{1}^{\dag}}{\partial t}\Bigg)-
\\&-\Bigg(e(\tilde{\psi}^{\dag}_{0}\tilde{\psi}_{1}+\tilde{\psi}^{\dag}_{1}\tilde{\psi}_{0})+\frac{e^2\phi_1}{mc^2}n_{0effective}\Bigg)
\end{align}
We use the following fourier representation for the  $\tilde{\psi}_1$ and for the scalar potential $\phi_1$
\begin{align}
\label{modfiedkgepp5}\phi_1(r,t)&=\hat{\phi}\text{exp}(iK\cdot r-i\Omega t)+\hat{\phi}^{\star}\text{exp}(-iK\cdot r+i\Omega t)
\\\label{modfiedkgepp15}\tilde{\psi}_1(r,t)&=\hat{\psi}_+\text{exp}(iK\cdot r-i\Omega) t+\hat{\psi}_{-}\text{exp}(-iK\cdot r+i\Omega t)
\end{align}
We separate all the  fourier modes in both the equations (\ref{modfiedkgepp5}) and (\ref{modfiedkgepp15}). Equations for each fourier mode produce exactly the same dispersion relation
\begin{align}
1+\chi_e-\chi_p=0
\end{align}
where
\begin{align}
\chi_e=\omega_{pe}^2\Bigg(\frac{4m^2_ec^4-\hbar^2 (\Omega^2- c^2K^2)  }{\hbar^2(\Omega^2- c^2K^2)^2-4 m^2_ec^4\Omega^2}\Bigg)
\end{align}
and
\begin{align}
\chi_p=\omega_{pp}^2\Bigg(\frac{4m^2_ec^4-\hbar^2 (\Omega^2- c^2K^2)  }{\hbar^2(\Omega^2- c^2K^2)^2-4 m^2_ec^4\Omega^2}\Bigg)
\end{align}
In the following Fig. we have plotted the dispersion curves for different values of the positron charge density.
\begin{figure}[!h]
  \centering
  \includegraphics[width=5 in]{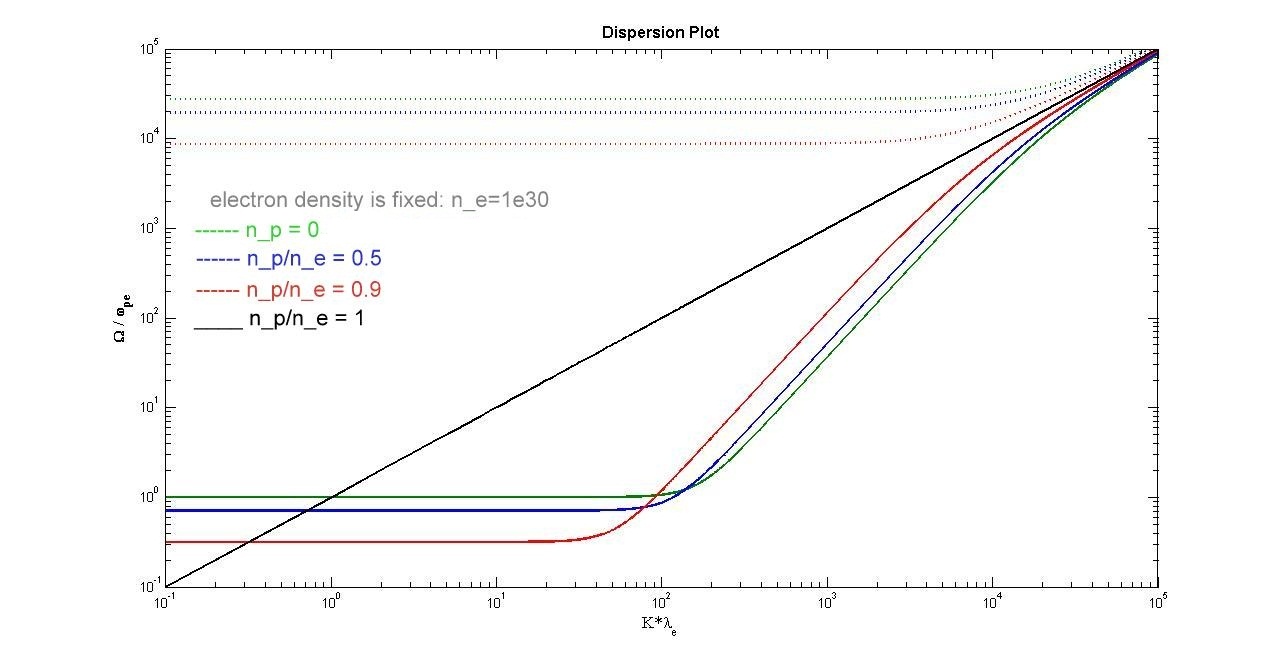}\\
  \caption{Dispersion Curves for Electron-Positron-Ion Plasma}\label{k}
\end{figure}
\newpage
\section{Conclusions}
In order to preserve causality  relativistic quantum plasma must contain antiparticles for dynamical particles.  It will be very interesting to further investigate the properties of these plasma especially in the limit when one of the particles or antiparticles become stationary and hence does not affect the dispersion of the electrostatic/electromagnetic waves.
\section{Acknowledgements}
We thank Higher Education Commission of Pakistan for providing financial support. Saqib Sharif would like to thank Bengt Eliasson for the useful discussion about their work.
\bibliographystyle{unsrt}
\bibliography{relativistic-quantum-plasma}

\begin{thebibliography}{10}

\bibitem{Bengt}
Bengt Eliasson and Padma~K. Shukla.
\newblock Relativistic laser-plasma interactions in the quantum regime.
\newblock {\em Phys. Rev. E 83, 046407 (2011)}.

\bibitem{Padma}
Bengt Eliasson and Padma~K. Shukla.
\newblock Relativistic x-ray free-electron lasers in the quantum regime.
\newblock {\em Phys. Rev. E 85, 065401(R) (2012)}.

\bibitem{Haas}
P.~K.~Shukla F.~Haas, B.~Eliasson.
\newblock Relativistic klein-gordon-maxwell multistream model for quantum
  plasmas.
\newblock {\em Phys. Rev. E 85, 056411 (2012)}.

\bibitem{Eliasson}
Bengt Eliasson and Padma~K. Shukla.
\newblock Nonlinear propagation of light in dirac matter.
\newblock {\em Phys. Rev. E 84, 036401 (2011)}.

\bibitem{WEINBERG}
Steven Weinberg.
\newblock {\em Gravitation and Cosmology: Principles and applications of the
  general theory of relativity}.
\newblock John Wiley and Sons, Inc. New York London Sydney Toronto, 1973.

\bibitem{Streater}
R.~F. Streater and A.~S. Wightman.
\newblock {\em PCT Spin and Statistics, and All That}.
\newblock W. A. Benjamin, New York, 1964.

\bibitem{0201503972}
Michael~E. Peskin and Dan~V. Schroeder.
\newblock {\em An Introduction To Quantum Field Theory (Frontiers in Physics)}.
\newblock {Westview Press}, 1995.

\bibitem{Villars}
Herman Feshbach and Felix Villars.
\newblock Elementary relativistic wave mechanics of spin 0 and spin 1/2
  particles.
\newblock {\em Reviews of modern Physics, Volume 30, Number 1, January, 1958}.

\bibitem{Victor}
Kenneth C.~Hines Victor~Kowalenko, Norman E.~Frankel.
\newblock Response theory of partilce-anti-particle plasmas.
\newblock {\em Physics Reports, Volume 126, Issue 3, p. 109-187}.

\bibitem{Quang}
Quang Ho-Kim and Xuan-Yem Pham.
\newblock {\em Elementary Particles and Their Interactions: Concepts and
  Phenomena}.
\newblock Springer-verlag, 1998.

\bibitem{Mark}
Mark Srednicki.
\newblock {\em Quantum Field Theory}.
\newblock {Cambridge University Press; 1 edition (February 5, 2007)}.

\end{thebibliography}

\end{document}